\documentclass[aps,floatfix,nofootinbib,onecolumn,superscriptaddress]{revtex4}
\pdfoutput=1

\usepackage{amsmath}
\usepackage{amssymb}
\usepackage{amsthm}
\usepackage{dcolumn}
\usepackage{epsfig}
\usepackage{graphics}
\usepackage{graphicx}
\usepackage{slashed,epsfig}
\usepackage{longtable}
\usepackage{color}

\usepackage{graphicx,amsmath,amssymb}
\usepackage{indentfirst}

\definecolor{navyblue}{rgb}{0,0.08,0.45}

\def\Dslash{\raise.15ex\hbox{/}\kern-.7em D}
\def\Pslash{\raise.15ex\hbox{/}\kern-.7em P}

 \newcommand{\threehalf}{{\frac{3}{2}}}

\newcommand{\beq}{\begin{equation}}
\newcommand{\enq}{\end{equation}}
\newcommand{\beqa}{\begin{eqnarray}}
\newcommand{\beqast}{\begin{eqnarray*}}
\newcommand{\enqa}{\end{eqnarray}}
\newcommand{\enqast}{\end{eqnarray*}}
\newcommand{\beml}{\begin{multline}}
\newcommand{\enml}{\end{multline}}
\newcommand{\nn}{\nonumber}
\newcommand{\req}[1]{(\ref{#1})}
\newcommand{\lb}{\label}
\newcommand{\rf}{\ref}
\newcommand{\ct}{\cite}

\newcommand{\cM}{\mathcal{M}}

\newcommand{\pa}{\partial}

\newcommand{\bec}{\begin{center}}
\newcommand{\enc}{\end{center}}
\newcommand{\beqo}{\begin{quote}}
\newcommand{\enqo}{\end{quote}}
\newcommand{\fn}{\footnote}
\newcommand{\half}{{\textstyle{\frac{1}{2}}}}
\newcommand{\quart}{{\textstyle{\frac{1}{4}}}}
\newcommand{\mbf}[1]{\mathbf{#1}}

\newcommand{\cA}{{\cal A}}
\newcommand{\cL}{{\cal L}}
\newcommand{\Z}{{\cal T}}
\newcommand{\cH}{{\cal H}}
\newcommand{\cD}{{\cal D}}
\newcommand{\cG}{{\cal G}}
\newcommand{\cN}{{\cal N}}
\newcommand{\cC}{{\cal C}}
\newcommand{\cF}{{\cal F}}
\newcommand{\cS}{{\cal S}}
\newcommand{\cR}{{\cal R}}
\newcommand{\bA}{\mbox{\bf A}}
\newcommand{\bB}{\mbox{\bf B}}
\newcommand{\bF}{\mbox{\bf F}}
\newcommand{\bG}{\mbox{\bf G}}
\newcommand{\bD}{\mbox{\bf D}}
\newcommand{\bU}{\mbox{\bf U}}
\newcommand{\bV}{\mbox{\bf V}}

\newcommand{\al}{\alpha}
\newcommand{\be}{\beta}
\newcommand{\ga}{\gamma}
\newcommand{\de}{\delta}
\newcommand{\ep}{\epsilon}
\newcommand{\ze}{\zeta}
\newcommand{\et}{\eta}
\newcommand{\io}{\iota}
\newcommand{\ka}{\kappa}
\newcommand{\la}{\lambda}
\newcommand{\rh}{\rho}
\newcommand{\si}{\sigma}
\newcommand{\varsi}{\varsigma}
\newcommand{\ta}{\tau}
\newcommand{\up}{\upsilon}
\newcommand{\ph}{\phi}
\newcommand{\vp}{\varphi}
\newcommand{\vph}{\varphi}
\newcommand{\ch}{\chi}
\newcommand{\ps}{\psi}
\newcommand{\om}{\omega}
\newcommand{\omis}{\omega_{!!!!!!!! \iota}}
\newcommand{\Al}{\Alpha}
\newcommand{\Be}{B}
\newcommand{\Ga}{\Gamma}
\newcommand{\De}{\Delta}
\newcommand{\Ep}{E}
\newcommand{\Ze}{Z}
\newcommand{\Et}{H}
\newcommand{\Th}{\Theta}
\newcommand{\Ka}{K}
\newcommand{\La}{\Lambda}
\newcommand{\Rh}{P}
\newcommand{\Si}{\Sigma}
\newcommand{\Ta}{T}
\newcommand{\Up}{\Upsilon}
\newcommand{\Ph}{\Phi}
\newcommand{\Ch}{X}
\newcommand{\Ps}{\Psi}
\newcommand{\Om}{\Omega}
\newcommand{\hPh}{\hat \Phi}

\begin{document}

\begin{flushright}
{
\small
SLAC--PUB-- 15937\\
\date{today}}
\end{flushright}

\vspace{40pt}

\title{Light-Front Holographic QCD and Color Confinement}

\author{Stanley~J.~Brodsky} \affiliation{SLAC National Accelerator Laboratory\\
Stanford University, Stanford, California 94309, USA}

\author{ Guy F. de T\'eramond} \affiliation{Universidad de Costa Rica, San Jos\'e, Costa Rica} 

\author{ Hans G\"unter Dosch} \affiliation {Institut f\"ur Theoretische Physik, Philosophenweg 16, Heidelberg, Germany}

\begin{abstract}
One of the most fundamental problems in Quantum Chromodynamics is to understand the origin of the mass scale which controls the range of color confinement and the hadronic spectrum.  For example, if one sets the Higgs couplings of quarks to zero, then no mass parameters appear in the QCD Lagrangian, and the theory is conformal at the classical level.   Nevertheless,  hadrons have a finite mass.  We show that a  mass gap and a fundamental color confinement scale arise when one extends the formalism  of de Alfaro, Fubini and Furlan
(dAFF)
 to frame-independent light-front Hamiltonian theory.    Remarkably,  the resulting light-front potential has a unique form of a harmonic oscillator in the light-front invariant 
impact variable 
if one requires that the action remains conformally invariant.   The result is  a single-variable relativistic equation of motion for  $q \bar q$ bound states, a ``Light-Front Shr\"odinger Equation", analogous to the nonrelativistic radial Schr\"odinger equation,
which incorporates color confinement and other essential spectroscopic and dynamical features of hadron physics, including a massless pion for zero quark mass and linear Regge trajectories with the same slope  in the radial quantum number  and orbital angular momentum.  
The same light-front  equations with the correct hadron spin dependence arise from the holographic mapping of  ``soft-wall model" modification of AdS$_5$ space with a specific dilaton profile.   
The corresponding light-front Dirac equation provides a dynamical and spectroscopic model of nucleons.  A fundamental mass parameter $\kappa$ appears, determining the hadron masses  and the length scale which underlies hadron structure.  Quark masses can be introduced to account for the spectrum of strange hadrons.
This 
Light-Front Holographic approach predicts not only hadron spectroscopy  successfully, but also hadron dynamics -- hadronic form factors, the QCD running coupling at small virtuality, the light-front wavefunctions of hadrons, $\rho$ electroproduction, distribution amplitudes, valence structure functions, etc. 
Thus  the combination of 
light-front dynamics, its holographic mapping to gravity  in a higher dimensional space
and the dAFF procedure  provides new  insight into the physics underlying color confinement, chiral invariance, and the QCD mass scale, among the most profound questions in physics.


\end{abstract}

\maketitle



\section{Introduction:  The Fundamental Mass Scale of QCD}

The QCD Lagrangian for massless quarks  is not only chirally invariant,  but also conformally invariant at the classical level.     The classical theory displays invariance under both scale  (dilatation) and special conformal transformations~\cite{Parisi:1972zy, Braun:2003rp}.   Nevertheless, the quantum theory which is built upon this conformal template displays color confinement and a mass gap, as well as asymptotic freedom.  One then confronts a  fundamental question: how does the mass scale  which determines the masses of the light-quark hadrons,  the range of color confinement, as well as the running of the coupling appear in QCD?

\begin{figure}[h]
\centering
\includegraphics[width=10.6cm]{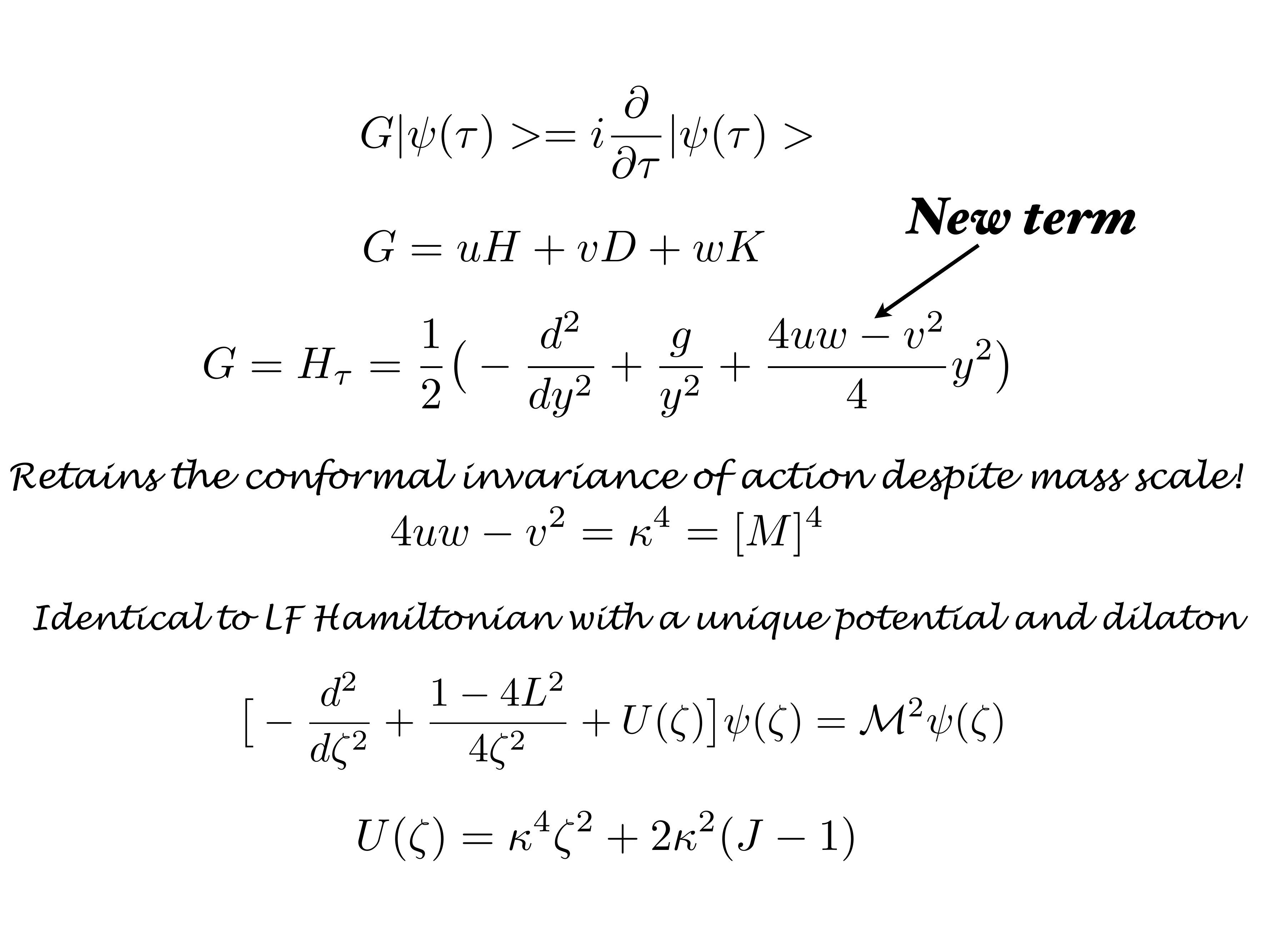}
 \caption{\small The application of the de Alfaro, Fubini and Furlan 
 procedure to the LF
 effective
 Hamiltonian.}
\label{dAFF}
\end{figure}

Important insight into the origin of the confinement scale and mass gap was given by de Alfaro, Fubini and Furlan (dAFF)~\cite{deAlfaro:1976je} in their 1976 paper on conformal invariance in one-dimensional quantum 
field theory.
They showed that, surprisingly, one can introduce a mass scale into the Hamiltonian and the equations of motion without destroying the conformal invariance of the action. 
In addition to the Hamiltonian $H$, the special conformal operator  $K$ and the dilatation operator $D$ are also conserved operators.
The crucial step is the introduction of a new Hamiltonian  $G$, a linear superposition of the generators $H$, $D$ and $K$, and consequently of a new evolution time variable $\tau$. 
A mass scale then arises without affecting the action. The 
new Hamiltonian acquires a harmonic oscillator potential as well as a new relative time variable $ \tau $ with finite range.  
See Fig. \ref{dAFF}. 
The form of the resulting potential is unique.    The mass parameter which enters the harmonic oscillator potential is in effect a place holder --  but it becomes fixed by one external measurement.

The dAFF 
procedure was originally developed 
in the context of 
nonrelativistic conformal quantum mechanics.
However, we have shown~\cite{Brodsky:2013ar} how to extend their method to the frame-independent light-front Hamiltonian~\cite{Brodsky:1997de} derived from the quantization of the QCD Lagrangian at fixed light-front time 
$x^+= x^0+x^3 = t + z/c$, the time along the front of a light wave -- Dirac's ``Front Form."

In principle, one can solve a quantum field theory such as QCD by computing  the eigenstates of the light front (LF) Hamiltonian, the operator 
$P^- = i \frac{\pa ~}{\pa x^+}$ of translations in $x^+$.
It is convenient to define  
the invariant LF Hamiltonian
 $H_{LF} = P_\mu P^\mu  =  P^- P^+ -  \vec{P}_\perp^2$ so that the basic Heisenberg equation is $H_{LF} \vert \phi \rangle  =  M^2 \vert \phi \rangle$~\cite{Brodsky:1997de}.  The eigenvalues of the light-front QCD Hamiltonian $H_{LF} $ predict the hadronic mass spectrum, and the corresponding eigensolutions provide the light-front wavefunctions which describe hadron structure.  The eigenstates  are defined at fixed 
 $x^+$ within the causal horizon, so that causality is maintained without normal-ordering.   In fact, light-front physics is a fully relativistic field theory, but its structure is similar to nonrelativistic atomic physics, and its bound-state equations can be formulated as relativistic Schr\"odinger-like equations at equal light-front time. 

Given the frame-independent light-front wavefunctions (LFWFs) $\psi_{n/H}$ , one can compute a large range of hadronic
observables, starting with form factors, structure functions, transverse momentum distributions generalized parton distributions, Wigner distributions, etc.
This formulation of nonperturbative QCD has the advantage that it is frame-independent, has no fermion-doubling, and acts in Minkowski space.   In the case of QCD (1+1) one can compute~\cite{Hornbostel:1988fb} 
the complete spectrum of meson and baryon states  to arbitrary precision for any number of colors, flavors, quark masses using the DLCQ method~\cite{Pauli:1985ps}. The mass of the mesons and baryon eigenstates at zero quark mass is determined  in QCD(1+1) in units of its dimensionful coupling.  

In principle, LF Hamiltonian theory provides a rigorous, relativistic, frame-independent framework for solving nonperturbative QCD and understanding the central problem of hadron physics -- color confinement.  However, in the case of 3+1 space-time, the QCD coupling is dimensionless, so the physical mechanism that sets the hadron mass scale for zero quark mass is not  apparent.  It is  important to note that the theory is intrinsically infrared divergent for on-shell quarks. For example, the ``H"  diagrams~\cite{Appelquist:1977tw}, the two-gluon ladder graphs, connecting a quark and antiquark becomes increasingly infrared logarithmically divergent~\cite{Smirnov:2009fh} as one adds rungs. However, this divergence is  eliminated if one postulates that the $q$ and $\bar q$ are always confined as color singlets with finite separation;   i.e., QCD(3+1) becomes infrared finite if one can introduce a confinement scale self-consistently.   It is thus compelling to apply the dAFF mechanism which can generate a mass scale and a confinement potential self-consistently without affecting the  conformal invariance of the action.

\section{The LF Schr\"odinger Equation}	

\begin{figure}[h]
\centering
\includegraphics[width=10.8cm]{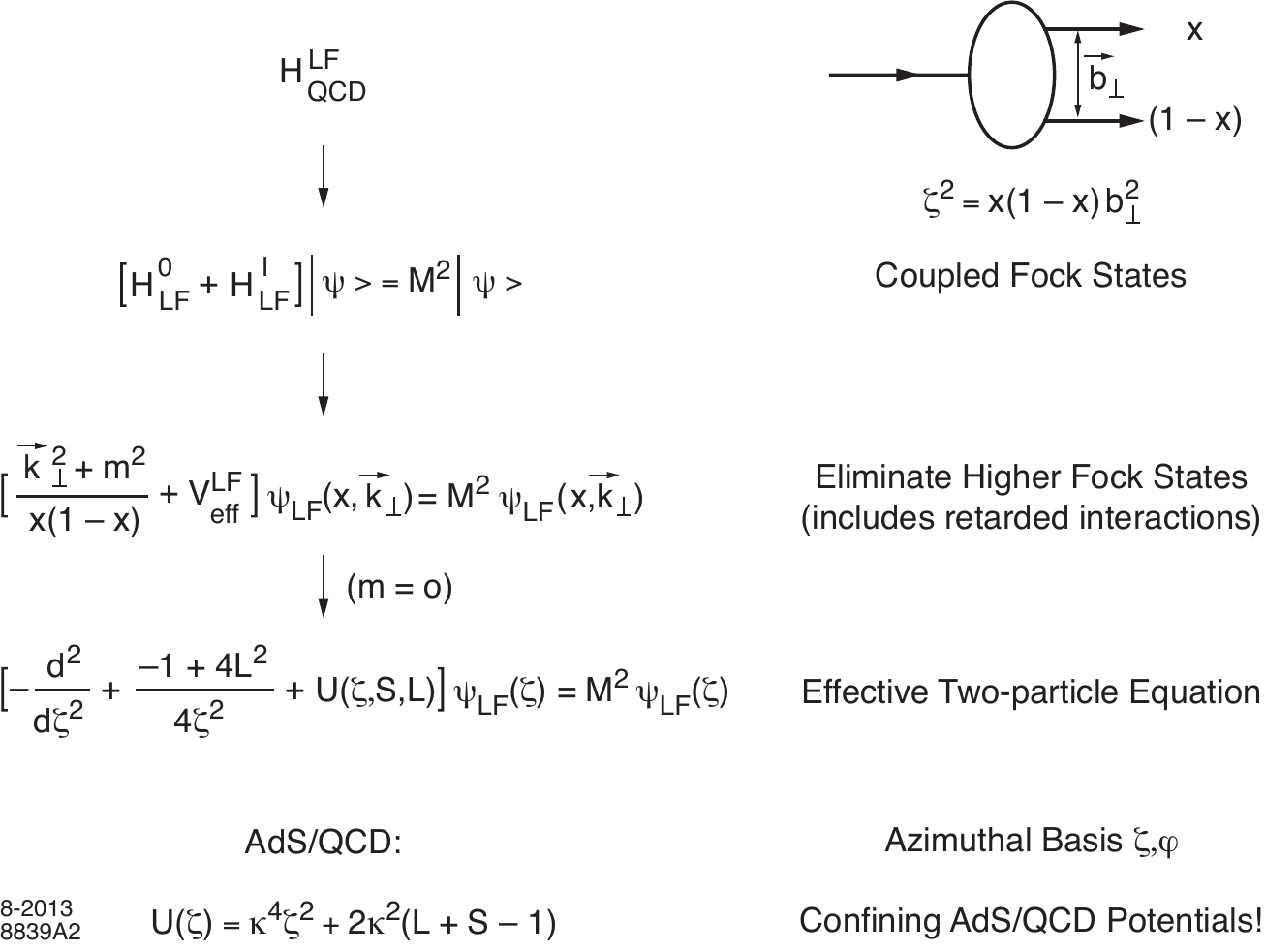}\hspace{0pt}
 \caption{\small Reduction of the QCD light-front Hamiltonian to an effective $q \bar q$ bound state equation. The harmonic oscillator form of $U(\zeta^2)$ is determined by the requirement that the action remain conformally invariant.  The additional constant term in the potential  $U(\zeta^2 )$ is determined from spin-$J$ representations on AdS$_5$
 embedding space. }  
\label{reduction}
\end{figure}

It is advantageous to reduce the full multiparticle eigenvalue problem of the LF Hamiltonian to an effective light-front Schr\"odinger equation  which acts on the valence sector LF wavefunctions of the lowest Fock State of a hadron~\cite{Pauli:1998tf}.  The central problem 
then becomes the derivation of the effective interaction 
$U_{\rm eff}$ which acts only on the valence sector of the theory and has, by definition, the same eigenvalue spectrum as the initial Hamiltonian problem.  In order to carry out this program one must systematically express the higher Fock components as functionals of the lower ones. This  method has the advantage that the Fock space is not truncated, and the symmetries of the Lagrangian are preserved~\cite{Pauli:1998tf}. See Fig. \ref{reduction}.
In the exact QCD theory the potential in the Light-Front Schr\"odinger equation (\ref{LFWE}) is determined from the two-particle irreducible (2PI) $ q \bar q \to q \bar q $ Green's function.  The elimination of the higher Fock states then
leads to an effective interaction $U\left(\zeta^2, J\right)$  for the valence $\vert q \bar q \rangle$ Fock state~\cite{Pauli:1998tf}.
A related approach for determining the valence light-front wavefunction and studying the effects of higher Fock states without truncation has been given in Ref.~\cite{Chabysheva:2011ed}.   

In the familiar case of nonrelativistic QED, one introduces angular coordinates $\theta$ and $\phi$ and the spherical harmonic basis to reduce the 3-dimensional equation to a one dimensional equation in the radial variable $r$. The kinetic energy acquires a term $\ell(\ell+1)/r^2$ for nonzero orbital angular momentum $\ell$. The dominant term in the effective potential is the Coulomb interaction.
In the case of the LF, the key radial variable is $\zeta^2 = b^2 x(1-x)$ the invariant separation between the quark and  antiquark where $x=k^+/P^+.$  The LF kinetic energy  -- which is  also the invariant mass squared $(p_q+ p_{\bar q})^2$ for a pair of massless quarks -- is 
${k^2_\perp\over x(1-x) } \to -{d^2\over d\zeta_\perp^2 }$.    
One then can introduce the azimuthal angle $\cal \phi$ and phase factor $\exp{(i L \cal \phi )} $ to obtain a one-dimensional LF Schr\"odinger equation with an extra kinetic energy term $(4 L^2 -1 )/\zeta^2$.
This is illustrated for a meson in Fig. \ref{reduction}.

In principle, the effective  potential $U_{\rm eff}$  in the single-variable  LF Schr\"odinger 
equation represents  the complete summation of interactions obtained from the Fock state reduction.  
It is an effective LF equation of motion acting on the lowest valence Fock state 
which encodes the fundamental conformal symmetry of the classical QCD Lagrangian.  We now apply the dAFF method to generate a mass term while retaining the conformal invariance of the action. The confining term in the LF potential can then have only one form: $\kappa^4 \zeta^2$, and the corresponding dilaton profile of the dual holographic AdS$_5$  model is uniquely determined to be $e^{\kappa^2 z^2} $. 

Thus light-front holography, together with the dAFF procedure, leads to an effective light-front Hamiltonian and relativistic frame-independent wave equation with a unique confining potential~\cite{deTeramond:2008ht} (See Fig. \ref{reduction})
\begin{equation} \label{LFWE}
\left[-\frac{d^2}{d\zeta^2}
- \frac{1 - 4L^2}{4\zeta^2} + U\left(\zeta^2, J\right) \right]
\phi_{n,J,L}(\zeta^2) = 
M^2 \phi_{n,J,L}(\zeta^2).
\end{equation}
This equation describes the spectrum of mesons as a function of $n$, the number of nodes in $\zeta$, the total angular momentum  $J$, which represent the maximum value of $\vert J^z \vert$, $J = \max \vert J^z \vert$,
and the internal orbital angular momentum of the constituents $L= \max \vert L^z\vert$.

\section{Confinement Scale and Uniqueness of the Effective Potential}

The effective theory should incorporate the fundamental conformal symmetry of the four-dimensional  classical QCD Lagrangian in the limit of massless quarks. To this end we study the invariance properties of a one-dimensional field theory under the full conformal group following the  dAFF construction of Hamiltonian operators described in Ref. ~\cite{deAlfaro:1976je}.

One starts with the one-dimensional action
\beq \label{S}
{ \cal{S}}= \half \int dt \left (\dot Q^2 - \frac{g}{Q^2} \right),
\enq
which  is invariant under conformal transformations in the 
variable $t$. In addition to the Hamiltonian  $H(Q, \dot Q) = \half \Big(\dot Q^2 + \frac{g}{Q ^2}\Big)$ there are two more invariants of motion for this field theory, namely the 
dilatation operator $D$ and $K$, corresponding to the special conformal transformations in $t$.  
Specifically, if one introduces the  new variable $\tau$ defined through 
$d\tau= d t/(u+v\,t + w\,t^2)$ and the  rescaled fields $q(\tau) = Q(t)/(u + v\, t + w \,t^2)^{1/2}$,
it then follows that the the operator
$G= u\,H+ v\,D + w\,K$
generates the quantum mechanical evolution in  $\tau$~\cite{deAlfaro:1976je}
 \beq    \label{EG}
G \vert \psi(\tau) \rangle = i \frac {d}{d \tau} \vert \psi(\tau)\rangle, \quad \quad 
 i \left[G, q(\tau) \right]  = \frac{d q(\tau)}{d \tau} .
 \enq

One can show explicitly~\cite{deAlfaro:1976je, Brodsky:2013ar} that  a confinement length scale appears in the action when one expresses the action \req{S} in terms of  the new time variable $\tau$ and the new fields $q(t)$, without affecting its conformal invariance. Furthermore, for $g \geq -1/4$ and $4\, u w -v^2 > 0$ the 
corresponding Hamiltonian 
\beq
G(q, \dot q)=  \half \Big( \dot q^2 + \frac{g}{q^2} +\frac{4\,uw - v^2}{4} q^2\Big),
\enq
is a compact operator.  Finally, we can transform back to the original field operator $Q(t)$ in \req{S}.  We find 
\beqa \label{HtauQ}
G(Q, \dot Q ) \! &= \! & \frac{1}{2} u \Big(\dot Q^2 + \frac{g}{Q^2} \Big)  - \frac{1}{4} v \Big( Q \dot Q + \dot Q Q\Big) + \frac{1}{2} w Q^2\\  \nn
            \! &= \! & u H+ v D + w K,
            \enqa
at $t=0$.  We thus recover the evolution operator $G =  u H + v D + w K$ which describes the evolution in the variable $\tau$, but expressed in terms of the original field $Q$.

The Schr\"odinger picture follows by identifying $Q \to x$ and $\dot Q \to -i {d\over dx}$. 
Then the evolution operator in the new time variable $\tau$ is
\beq \label{Htaux} 
G = {1\over 2} u \Big(-{d^2\over dx^2}  + {g\over x^2} \Big) + {i\over 4} v \Big(x {d\over dx} + {d\over dx}x \Big) +{1\over 2}wx^2
\enq

We now compare   the Hamiltonian \req{Htaux}  with the light-front wave equation \req{LFWE} and identify the variable $x$ with the light-front invariant variable $\zeta$.   We then choose $u=2, \; v=0$ and identify the dimensionless constant $g$ with the LF orbital angular momentum $g=L^2-1/4$  to reproduce the light-front kinematics. Identifying  $w = 2 \lambda^2$ then fixes the confining light-front  potential to a quadratic $\la^2 \, \zeta^2$ form.
The mass scale brought in via $w = 2\kappa^4$ then generates the confining mass scale $\kappa$.

\section{Light-Front Holography and AdS/QCD}

Anti-de Sitter space in five dimensions plays a special role in elementary particle physics since it provides an exact geometrical representation of the conformal group.  In fact, gravity  in AdS$_5$ space is holographically dual to frame-independent light-front Hamiltonian theory.   Light-front holography also leads to a precise relation between the bound-state amplitudes in the fifth dimension $z$ of AdS space and the variable $\zeta$, the argument of the boost-invariant light-front wavefunctions describing the internal structure of hadrons in physical space-time.      The variable $z$ of AdS space is identified with the LF   boost-invariant transverse-impact variable $\zeta$~\cite{Brodsky:2006uqa}, 
thus giving the holographic variable a precise definition in LF QCD~\cite{deTeramond:2008ht, Brodsky:2006uqa}.
For a two-parton bound state $\zeta^2 = x(1-x) b^{\,2}_\perp$.
Remarkably, the light-front Hamiltonian equations of motion~\cite{deTeramond:2008ht}  then have a structure which matches exactly the eigenmode equations in AdS space. This makes a direct connection of QCD with AdS methods possible:  the same light-front  equations emerge from AdS/QCD and light-front holography, the holographic mapping between the 
holographic variable $z$ in five-dimensional anti-deSitter space and $\zeta$ in 3+1 QCD.   

The ``soft-wall" modification of the AdS$_5$ metric through the dilaton factor $e^{\pm\kappa^2 z^2}$ which multiplies the $AdS_5$ 
Lagrangian provides a simple analytic way to break its scale invariance. In fact, if one modifies the conformal AdS action by the soft-wall dilaton $\exp(+\kappa^2 z^2)$, the resulting equation of motion is identical to the LF equation derived from the dAFF 
procedure; in addition, a constant term $2 \kappa^2 (J-1)$ appears in the LF potential when one uses AdS$_5$ to represent a bound state with  arbitrary
spin  $J$~\cite{deTeramond:2013it} . The positive exponential insures stability of strings in the corresponding five-dimensional gravity theory~\cite{Klebanov:2009zz}

\begin{figure}[h]
\centering
\includegraphics[width=5.0cm]{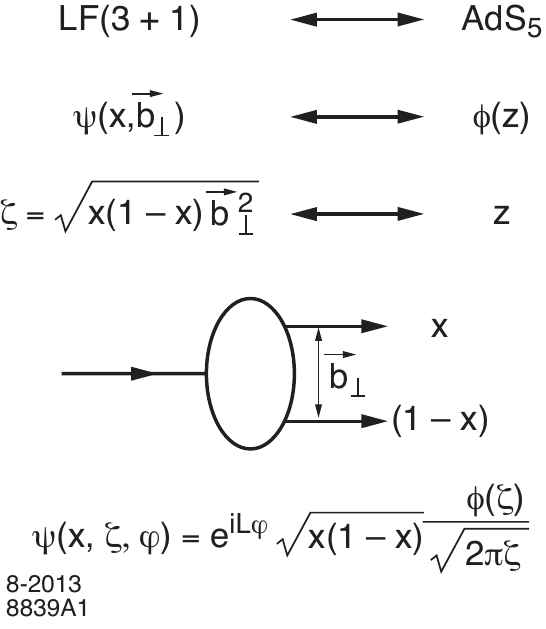}\hspace{0pt}
 \caption{\small The dictionary underlying light-front holography~\cite{deTeramond:2008ht}, the correspondence between AdS$_5 $ and QCD(3+1).}
\label{LFHolography}
\end{figure} 

\begin{figure}[h]
\centering
\includegraphics[width=5.8cm]{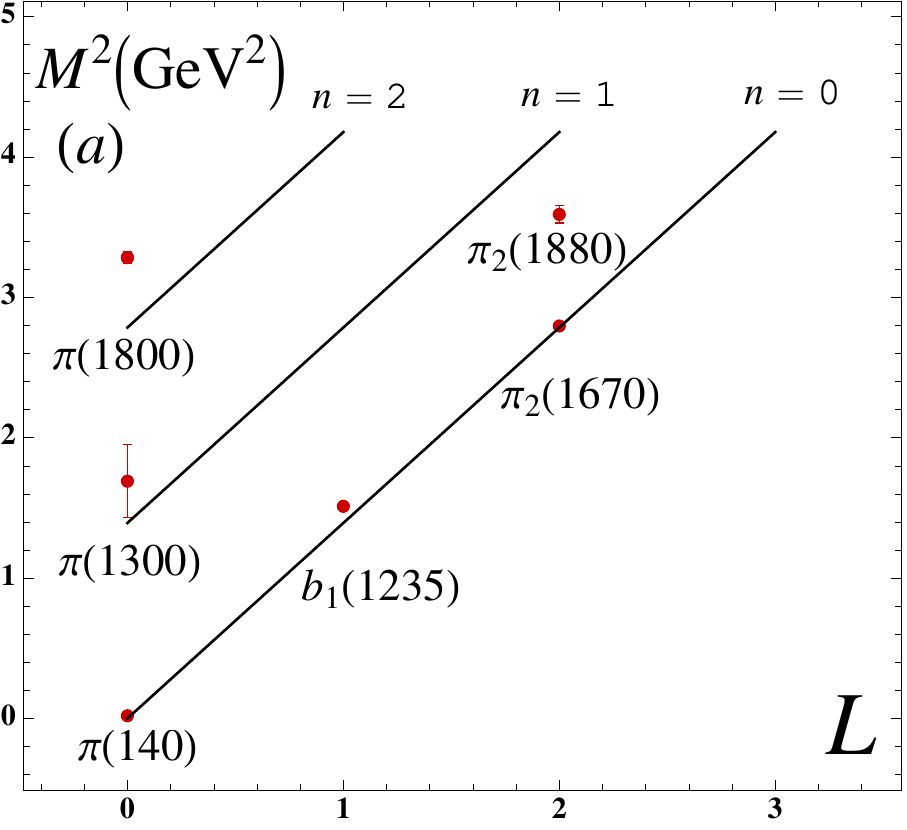}  \hspace{20pt}
\includegraphics[width=5.8cm]{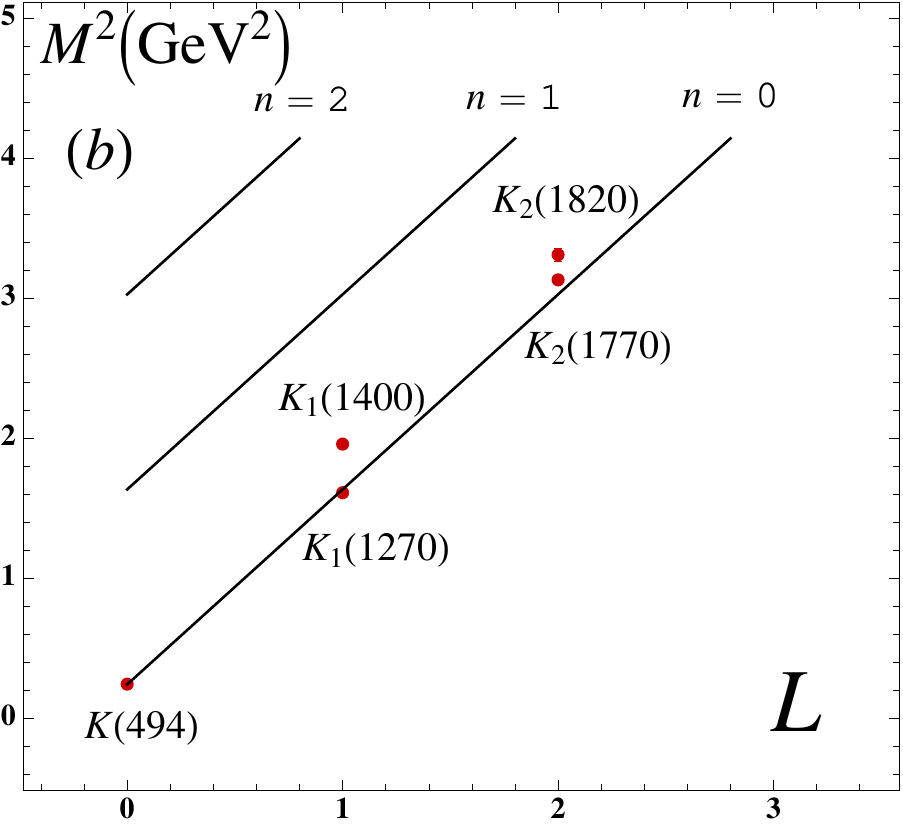}  
\includegraphics[width=5.8cm]{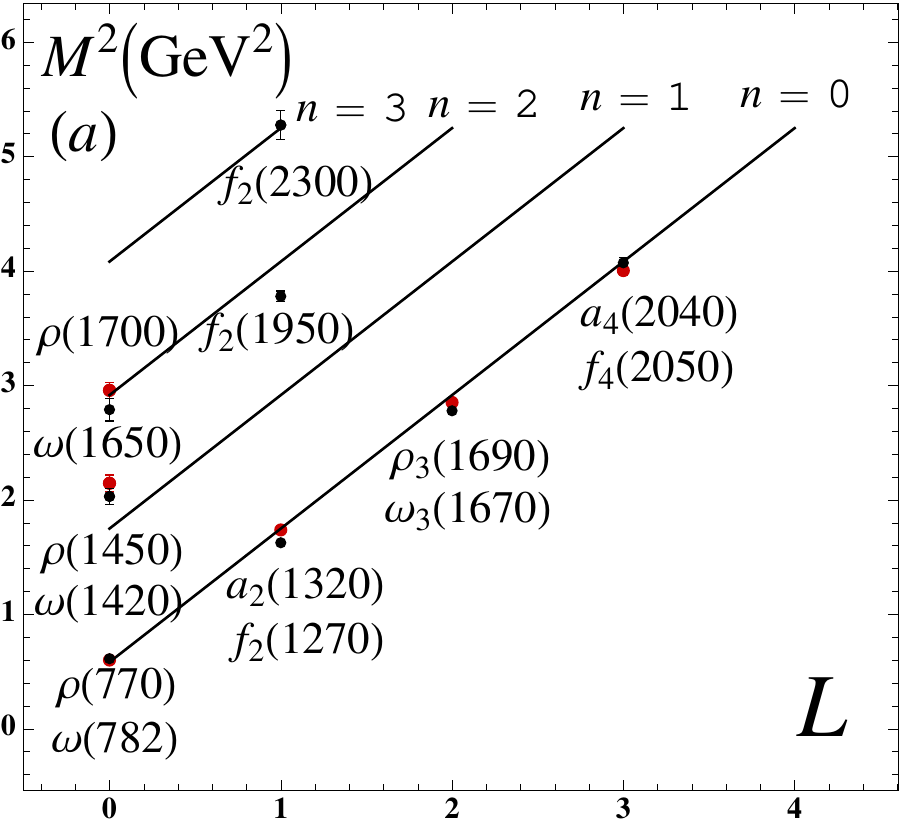}  \hspace{20pt}
\includegraphics[width=5.8cm]{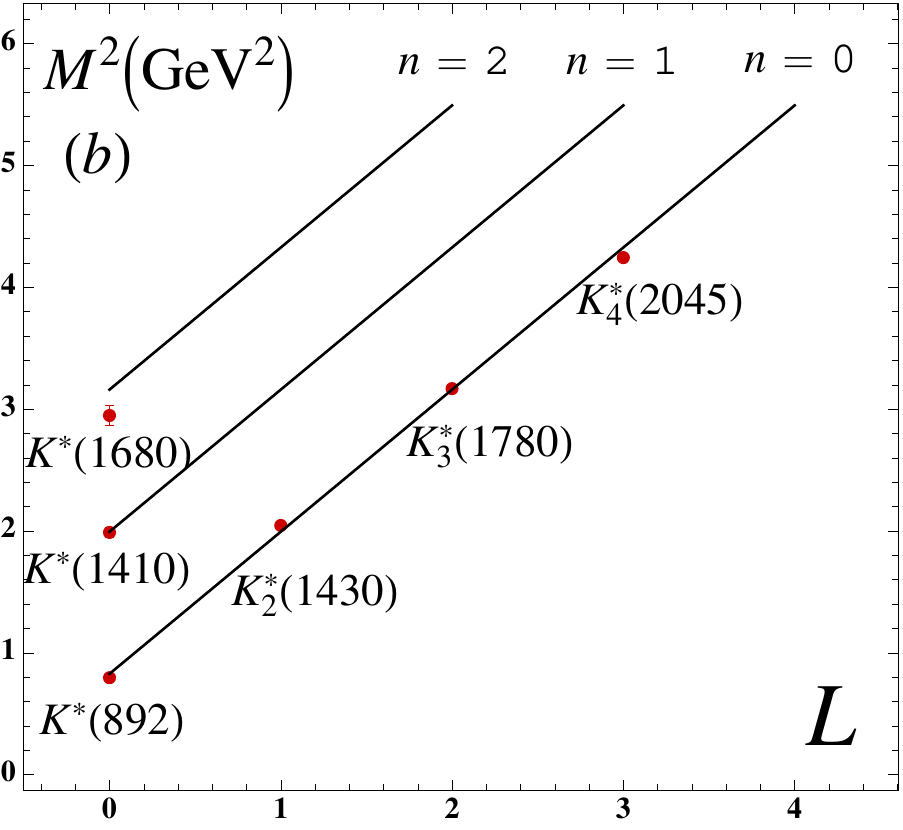}
\caption{Orbital and radial excitations for $\ka = 0.59$ GeV  (pseudoscalar) and  0.54 GeV (vector mesons)}
\label{pionspec}
\end{figure}

Light-Front Holography is illustrated in Fig. \ref{LFHolography}. The dilaton profile $\exp{\left(\pm \kappa^2 z^2\right)}$  
leads to linear Regge trajectories~\cite{Karch:2006pv}. More specifically,
the dilaton $\exp{\left(\kappa^2 z^2\right)}$ in AdS space, together with the identification of $z$ and $\zeta$ leads to the effective potential
$U(\zeta^2,J) =   \kappa^4 \zeta^2 + 2 \kappa^2(J - 1)$  
and to eigenvalues
$M_{n, J, L}^2 = 4 \kappa^2 \left(n + \frac{J+L}{2} \right)$~\cite{deTeramond:2013it, Gutsche:2011vb}.
We thus obtain a nonperturbative relativistic light-front quantum mechanical wave equation for mesons which incorporates color confinement and other essential spectroscopic and dynamical features of hadron physics, including a massless pion for zero quark mass and linear Regge trajectories $M^2(n, L, S) = 4\kappa^2( n+L +S/2)$ with the same slope  in the 
radial quantum number $n$ and orbital angular momentum $L$.    
If one changes the power in the dilaton profile to  $z^p$, the mass of the $J=L=n=0$  pion is 
 zero in the chiral limit only for 
 the
 case $p=2$. To show this, one can use the stationarity of bound-state energies with respect to variation of parameters.  Predictions  for the spectrum of  light pseudoscalar and vector meson  states are  compared with experimental data in Fig. \ref{pionspec}.  The separate dependence on $J=L+S$ and $L$ leads to a  mass ratio of the $\rho$ and the $a_1$ mesons which coincides with the result of the Weinberg sum rules~\cite{Weinberg:1967kj}. 
The predictions can also be extended to other light hadron families, as for example the strange vector meson $K$ and $K^*$  families,~\cite{Dosch:2014wxa} which are also included in Fig. \ref{LFHolography}.  In fact, to first approximation  one expects that the effective potential, and thus the transverse dynamics, are unchanged to first order in the quark masses~\cite{Brodsky:2008pg}.

Recently we have derived wave equations for hadrons with arbitrary spin starting from an effective action in  AdS space~\cite{deTeramond:2013it}.    An essential element is the mapping of the higher-dimensional equations  to the LF Hamiltonian equation  found in Ref.~\cite {deTeramond:2008ht}.  This procedure allows a clear distinction between the kinematical and dynamical aspects of the LF holographic approach to hadron physics.  Accordingly, the non-trivial geometry of pure AdS space encodes the kinematics,  and the additional deformations of AdS encode the dynamics, including confinement~\cite{deTeramond:2013it}, and determine the form of the LF effective potential from the precise holographic mapping to light-front physics~\cite{deTeramond:2008ht, deTeramond:2013it}.  One finds  from the dilaton-modified AdS action the LF potential
\beq \label{U}
U(\ze, J) = \frac{1}{2}\vp''(\ze) +\frac{1}{4} \vp'(\ze)^2  + \frac{2J - 3}{2 \zeta} \vp'(\ze) .
\enq
The correspondence between the LF and AdS equations  thus determines the effective confining interaction $U_{\rm eff}$ in terms of the infrared behavior of AdS space and gives the holographic variable $z$ a kinematical interpretation. The identification of the orbital angular momentum with the mass term in the AdS action $\mu^2R^2= L^2-(J-2)^2$
is also a key element of the description of the internal structure of hadrons using holographic principles. 
 he choice of the dilaton profile $\varphi(z) = \ka z^2$  thus follows  from the requirements of conformal invariance. 

The triple-complementary connection of  (a)  AdS space,  (b) its LF holographic dual, 
and (c) the dAFF relation to the algebra of the conformal group in one dimension, is  thus characterized by a unique quadratic confinement LF potential.
Identical light-front  equations arise from the holographic mapping 
of the soft-wall model modification of AdS$_5$ space with a unique dilaton profile  to QCD (3+1) at fixed light-front time~\cite{deTeramond:2013it}.  Light-front holography thus provides a precise relation between the bound-state amplitudes in the fifth dimension of AdS space and the boost-invariant light-front wavefunctions describing the internal structure of hadrons in physical space-time. 
The AdS approach, however,  goes beyond the purely group theoretical considerations of dAFF, since 
features such as the masslessness of the pion and the separate dependence on $J$ and $L$ are a consequence of the  potential $U(\zeta^2,J)$ derived from the duality with AdS for general high-spin representations.~\cite{deTeramond:2013it}

The effective interaction $U(\zeta^2,J)$
is instantaneous in LF time and acts on the lowest state of the LF Hamiltonian.  The resulting LF Schr\"odinger equation describes the spectrum of mesons as a function of $n$, the number of nodes in $\zeta^2$,
 the internal orbital angular momentum $L = L^z$, and the total angular momentum $J=J^z$,
with $J^z = L^z + S^z$  the sum of the  orbital angular momentum of the constituents and their internal spins.
The  ${\rm SO(2)}$ Casimir  $L^2$  corresponds to  the group of rotations in the transverse LF plane.

The result from both AdS/QCD and the application of dAFF is a nonperturbative relativistic light-front quantum mechanical wave equation for mesons which incorporates color confinement and other essential spectroscopic and dynamical features of hadron physics.
Only one mass parameter $\kappa$ appears, determining the hadron masses such as the $\rho$ and proton,  and  hadron sizes.
The  LF Holography approach successfully predicts not only hadron spectroscopy, but also hadron dynamics -- hadron form factors, the QCD running coupling at small $Q^2$, LFWFs, $\rho$ electroproduction, distribution amplitudes, valence structure functions -- 
 in terms of one parameter, the mass scale.

\begin{figure}[h]
\centering
\includegraphics[width=10.0cm]{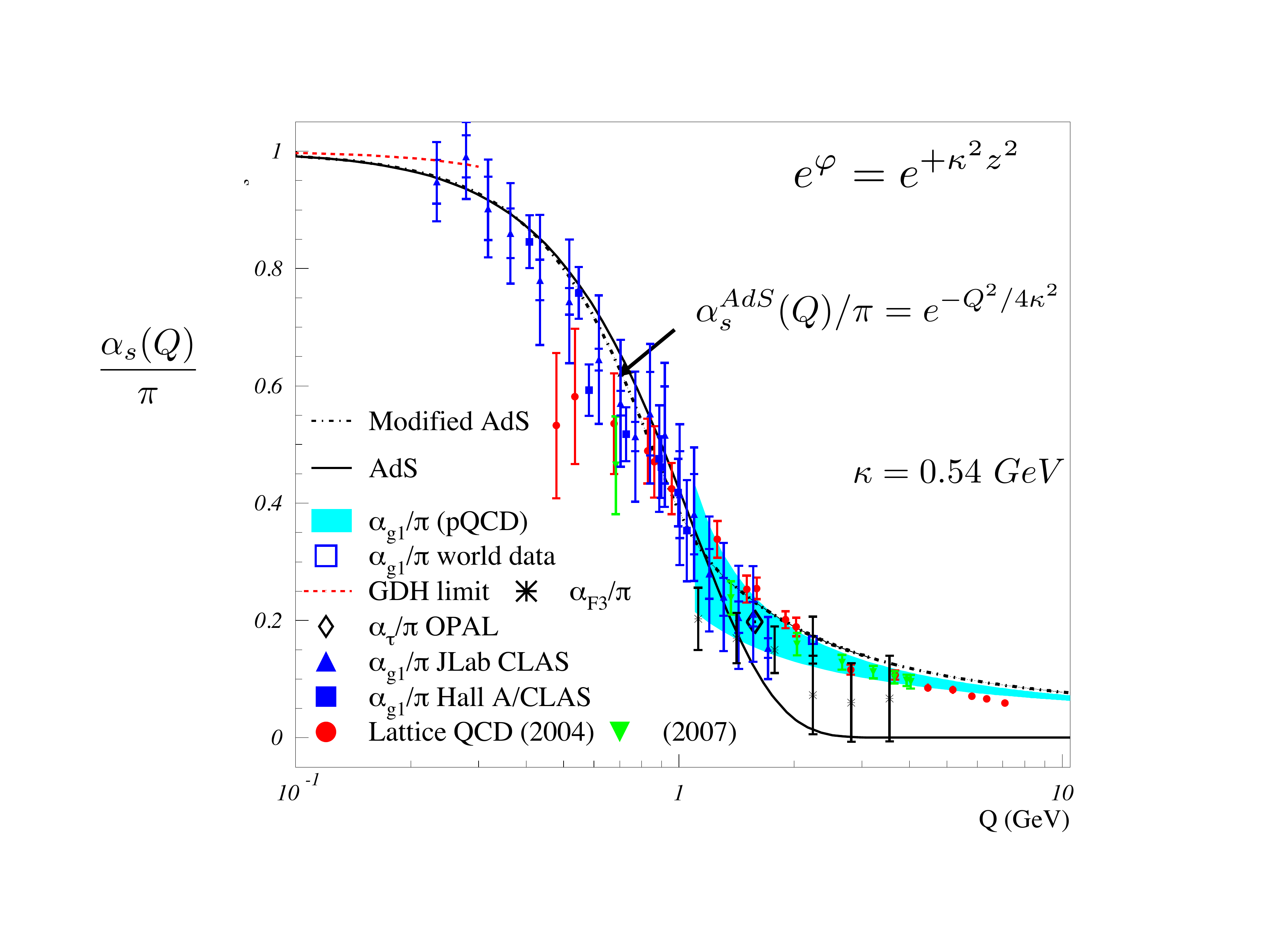} 
 \caption{\small Light-front holographic results for the QCD running coupling from Ref. ~\cite{Brodsky:2010ur} normalized to $\alpha_s(0)/ \pi =1.$  
 The result is analytic, defined at all scales and exhibits an infrared fixed point.}
 \label{alphas}
\end{figure}

The QCD mass scale $\kappa$ in units of GeV can be determined by one measurement; e.g., the pion decay constant $f_\pi.$  All other masses and size parameters are then predicted. The running of the QCD coupling  is predicted in the infrared region for $Q^2 <  4 \kappa^2$  to have the  form $\alpha_s(Q^2) \propto \exp{\left(-Q^2/4 \kappa^2\right)}$. As shown in Fig. \ref{alphas}, the result agrees with the shape of the effective charge defined from the Bjorken sum rule~\cite{Brodsky:2010ur}, displaying an infrared fixed point.  In the nonperturbative domain soft gluons are in effect sublimated into the effective confining potential. Above this region, hard-gluon exchange becomes important, leading to asymptotic freedom.  
The scale $\Lambda$ entering the evolution of the  perturbative QCD running constant  in a given renormalization scheme such as $\Lambda_{\overline MS}$ could be determined in terms of the primary scheme-independent scale $\kappa$ 

The new time variable $\tau$ that appears in the dAFF procedure is related to the variable $t$ for the case  $u w >0, \,v=0$   by 
$\tau =\frac{1}{\sqrt{u\,w}} \arctan\left(\sqrt{\frac{w}{u}} t\right),$
{\it i.e.}, $\tau$ has only a limited range. The finite range of invariant LF time $\tau=x^+/P^+$ can be interpreted as a feature of the internal frame-independent LF  time difference between the confined constituents in a bound state. For example, in the collision of two mesons, it would allow one to compute the LF time difference between the two possible quark-quark collisions~\cite{Brodsky:2013ar}.

The LF Schr\"odinger equation is the relativistic frame-independent front-form analog of the non-relativistic radial Schr\"odinger equation for muonium  and other hydrogenic atoms in presence of an instantaneous Coulomb potential. The LF harmonic oscillator potential could in fact emerge from the exact QCD formulation when one includes contributions from the 
effective potential $U$ which are due to the exchange of two connected gluons; {\it i.e.}, ``H'' diagrams~\cite{Appelquist:1977tw}.
We notice that $U$ becomes complex for an excited state since a denominator can vanish; this gives a complex eigenvalue and the decay width.  The multi-gluon exchange diagrams also could be connected to the Isgur-Paton flux-tube model of confinement; the collision of flux tubes could give rise to the ridge phenomena observed in high energy $pp$ collisions at RHIC~\cite{Bjorken:2013boa}. The quadratic form of the effective light-front potential can also be derived by implying the Ehrenfest principle to the front form formalism. 
Further details are given in Ref.~\cite{Glazek:2013jba}

\begin{figure}[h]
\centering
\includegraphics[width=8.4cm]{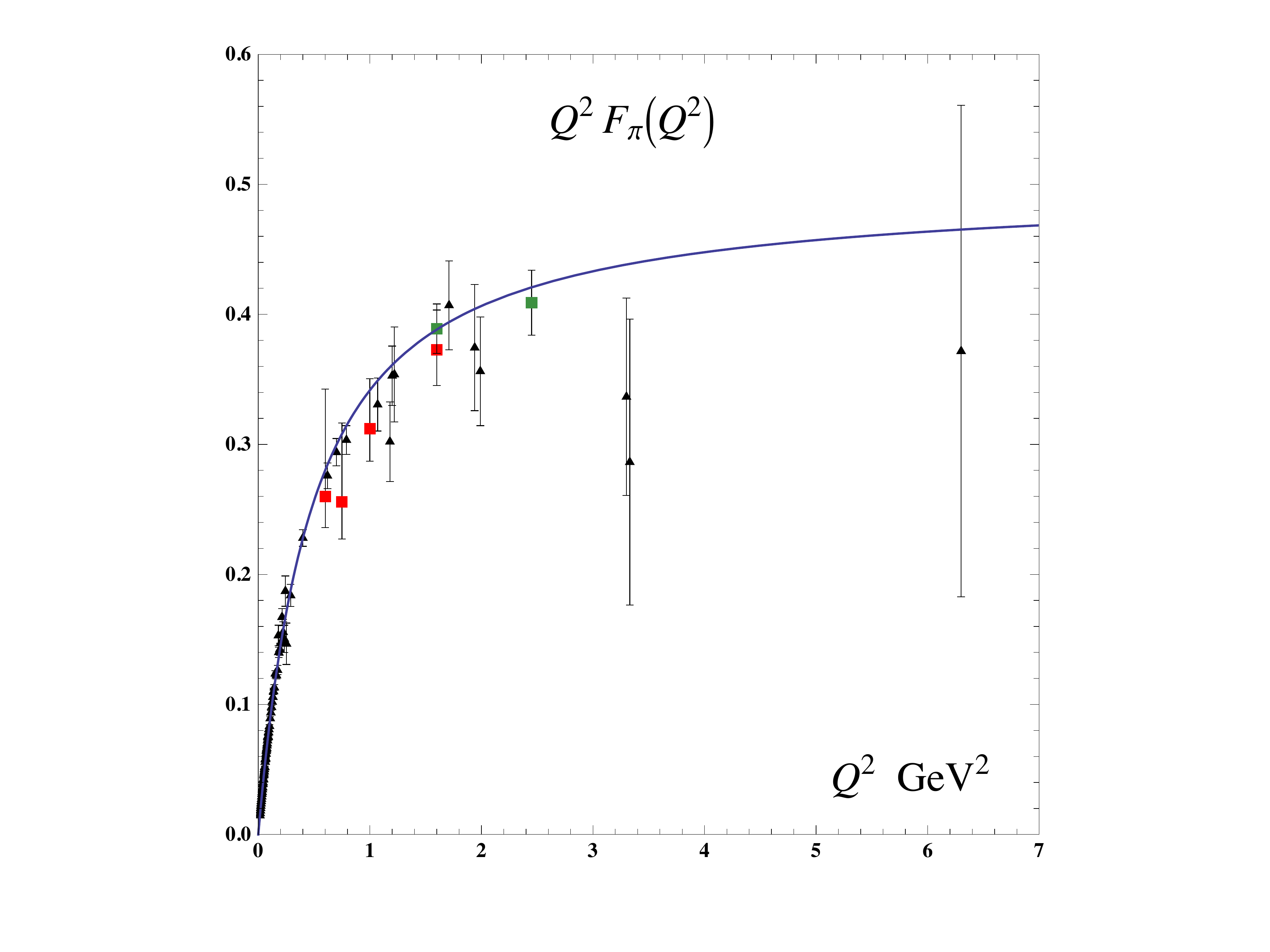} \hspace{0pt}
 \caption{\small Light-front holographic prediction for the space-like pion form factor.  }
\label{pionff}
\end{figure} 

We have also applied  LF holography to space-like  (see Fig.  \ref{pionff}) and time-like meson form factors~\cite{deTeramond:2012rt}, as well as transition amplitudes such as 
$\gamma^* \gamma \to \pi^0$~\cite{Brodsky:2011xx}, $\gamma^* N \to N^*$, all based on essentially the single mass scale parameter $\kappa.$   The timelike pion and nucleon form factors have poles at the $\rho, \rho^\prime$, etc, when one uses the dressed LF current computed from the behavior of the 
non-normalizable solution for the vector amplitude in a
dilaton-modified AdS space.  
Holographic QCD  incorporates important elements for the study of hadronic form factors which encompasses perturbative and nonperturbative elements, such as the connection between the twist of the hadron to the fall-off of its current matrix elements for large $q^2$, and essential aspects of vector meson dominance which are relevant at lower energies.  However, the description of form factors and transition amplitudes is still far from being fully understood in the light-front holographic  approach described here, and some {\it ad hoc }elements have to be introduced for a meaningful comparison with data.~\cite{deTeramond:2012rt}

\begin{figure}[h]
\includegraphics[width=13.8cm]{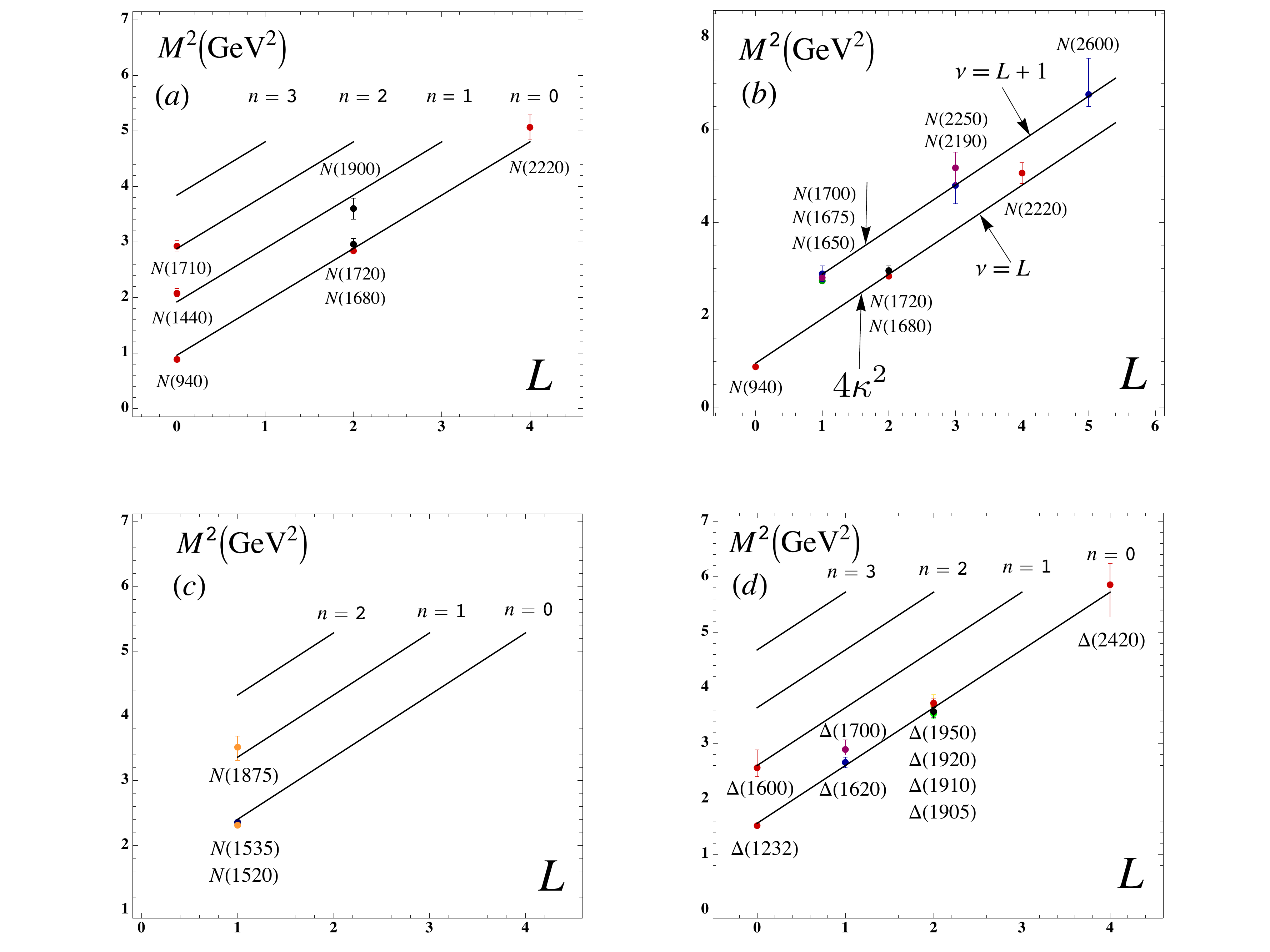}
 \caption{\small Light front holographic predictions of the light-front Dirac equation for the nucleon spectrum.    
 Positive-parity spin-$\half$ nucleons (a) and spectrum gap between the negative-parity spin-$\threehalf$ and the positive-parity spin-$\half$ nucleons families (b). Minus parity $N$ ({c}) and plus and minus parity $\Delta$  families (d), for $\ka = 0.49$ GeV (nucleons) and  0.51 GeV (Deltas). 
 All confirmed positive and negative-parity resonances from PDG 2014 are well accounted using the procedure described in \cite{deTeramond:2014yga}.}
\label{nucleonspect}
\end{figure} 

\begin{figure}[h]
\centering
 \includegraphics[width=6.0cm]{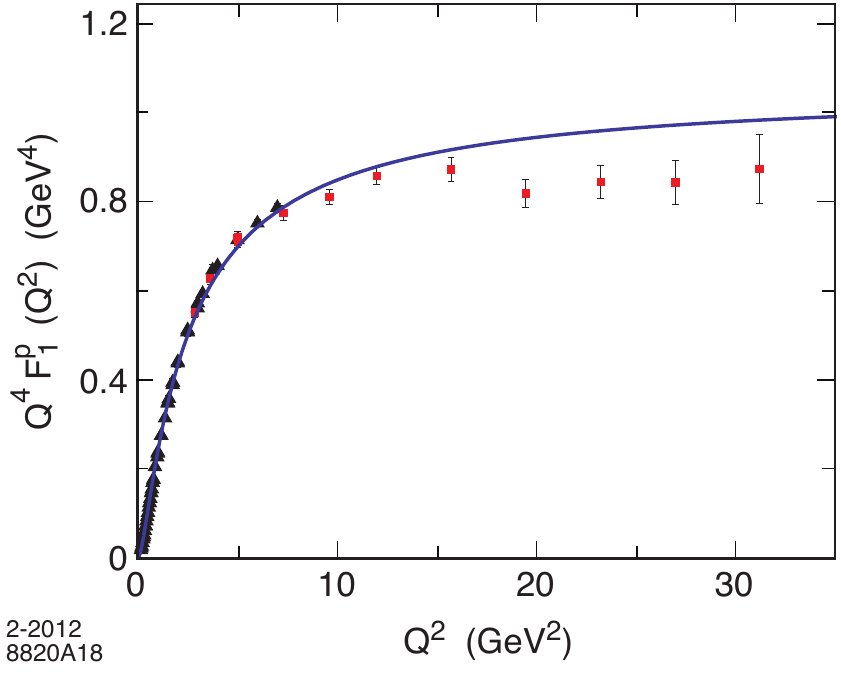}   \hspace{5pt}
\includegraphics[width=6.0cm]{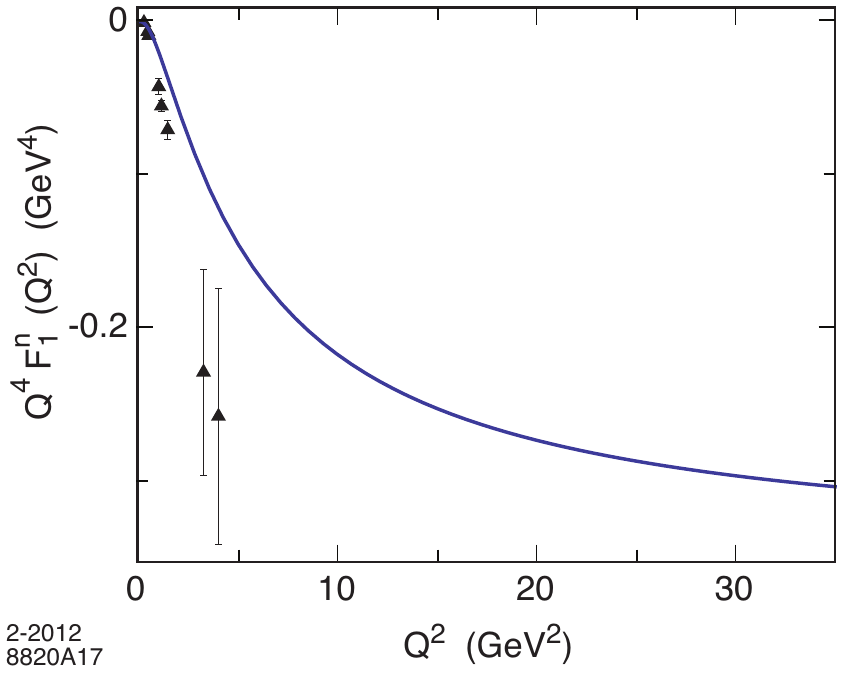}
 \includegraphics[width=6.0cm]{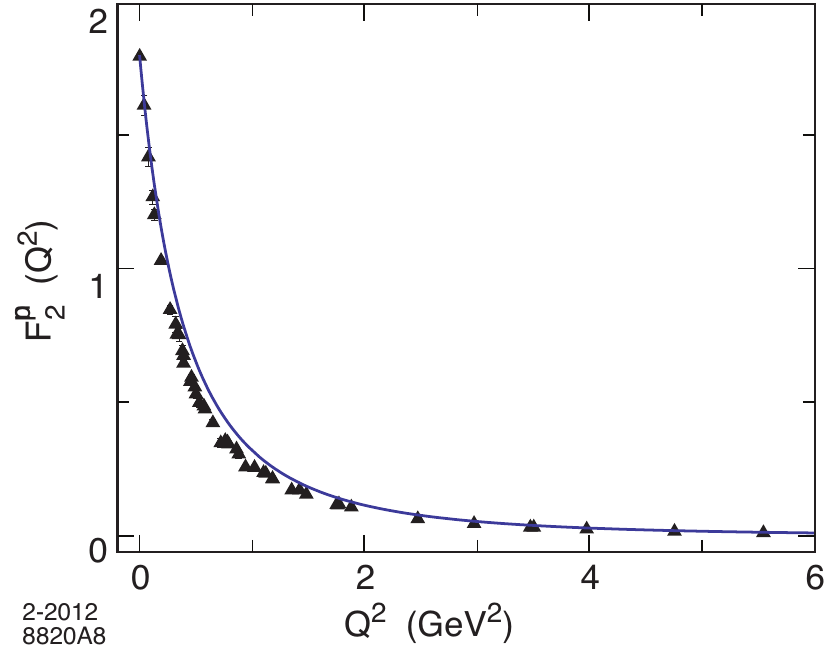}   \hspace{5pt}
\includegraphics[width=6.0cm]{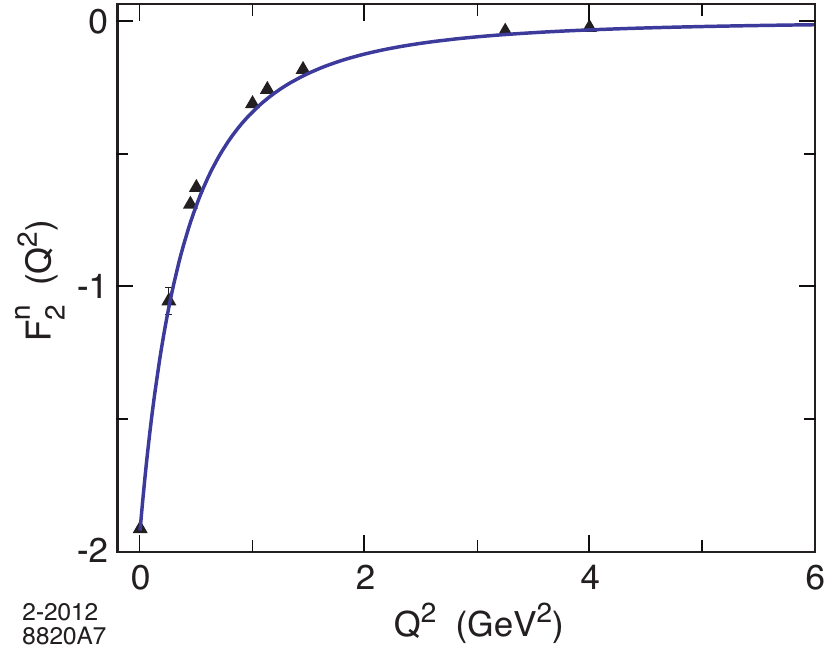}
\caption{Light-front holographic predictions for the nucleon form factors normalized to their static values.  }
\label{nucleonffs}
\end{figure}

The corresponding light-front Dirac equation provides a dynamical and spectroscopic model of nucleons. See Fig. \ref{nucleonspect}. Many other applications have been presented in the literature, including recent results by Forshaw and Sandapen~\cite{Forshaw:2012im} for diffractive  $\rho$ electroproduction which are  based on the light-front holographic prediction for the longitudinal $\rho$ LFWF.  Other recent applications include predictions for  generalized parton distributions (GPDs)~\cite{Vega:2010ns}, and a model for nucleon and flavor form factors~\cite{Chakrabarti:2013dda}. The 
light-front holographic
predictions~\cite{deTeramond:2012rt} for the 
nucleon form factors
are shown in Fig. 
\ref{nucleonffs}.
A detailed discussion of the light meson and baryon spectrum as well as  the elastic and transition form factors of the light hadrons using LF holographic methods is given in Ref.~\cite{deTeramond:2012rt}.

It should be noted that there is no contradiction between the confining long-distance  linear potential $\sigma r$ that appears in the effective nonrelativistic Schr\"odinger equation which binds heavy quarks   such as the Cornell potential, and the quadratic form of the harmonic oscillator potential $\kappa^4 \zeta^2$ which confines light quarks.   One can show~\cite{Trawinski:2014msa} that an instantaneous linear potential $V_{eff} $ in the instant form implies a quadratic potential 
$U_{\rm eff}$ in the front form at large $q \bar q$ separation:
\beq
U_{\rm eff} = V^2_{\rm eff} + 2 \sqrt{\vec p^2+ m_q^2} \, V_{\rm eff} + 2  V_{\rm eff} \sqrt{\vec p^2+ m_q^2}.
\enq
This result follows from the comparison of the invariant mass in the instant form in the center of mass system $\vec P=0$ with the invariant mass in the front form in the constituent rest frame
$\vec p_q + \vec p_{\bar q}=0.$  Furthermore, the ranges of the quark and antiquark are numerically similar when one uses the WKB method to determine their maximum separation. 
One can also show~\cite{Trawinski:2014msa} how the two-dimensional front-form harmonic oscillator potential for massless quarks takes on a three-dimensional form when the quarks have mass. The effective confining potential can  be modified~\cite{Trawinski:2014msa} by replacing $\zeta^2_\perp \to \zeta^2_\perp +\zeta_3^2$ where $\zeta_3= i {\partial \over \partial k_3}$ is conjugate to $k_3 = {2m(x-1/2)\over \sqrt{x(1-x)}}.$  The LF kinetic energy operator  for massive fields is $\sum_i \frac{k^2_{\perp i} + m_i^2}{ x_i}$ summed over the constituents.

\section{Chiral Symmetry Breaking and the Light-Front Vacuum}

It is conventional to define the vacuum in quantum field theory as the lowest energy eigenstate of the instant-form Hamiltonian.  Such an eigenstate is defined at a single time $t$ over all space $\vec x$.  It is thus acausal and frame-dependent. The instant-form vacuum thus must be normal-ordered in order to avoid violations of causality when computing correlators and other matrix elements. In contrast, in the front form, the vacuum state is defined as the  eigenstate of lowest invariant mass 
$M^2$. It is defined at fixed light-front time $x^+ = x^0 + x^3$ over all $x^-= x^0 - x^3$ and $\vec x_\perp$, the extent of space that can be observed within the speed of light.   It is frame-independent and only requires information within the causal horizon.

Since all particles have positive $k^+= k^0 + k^z > 0$ and $+$ momentum is conserved in the front form, the usual vacuum bubbles are kinematically forbidden in the front form.  In fact the LF vacuum for QED, QCD, and even the Higgs Standard Model is trivial up to possible zero modes -- backgrounds with zero four-momentum. In this sense it is already normal-ordered. The frame-independent causal front-form vacuum is a good match to the ``void" -- the observed universe without luminous matter.    Thus it is natural in the front form to obtain zero cosmological constant from quantum field theory~\cite{Brodsky:2009zd,Brodsky:2012ku}.     In the case of the Higgs theory, the usual Higgs vacuum expectation value is replaced by a classical $k^+=0$  background zero-mode field  which is not sensed by the energy momentum tensor~\cite{Srivastava:2002mw}.  The phenomenology of the Higgs theory is unchanged.

There are thus no quark or gluon vacuum condensates in the LF vacuum-- as first noted by Casher and Susskind~\cite{Casher:1974xd}; the corresponding physics is contained within the LFWFs themselves~\cite{Brodsky:2009zd,Brodsky:2010xf,Chang:2013pq,Chang:2013epa,Glazek:2011vg}, thus eliminating a major contribution to the cosmological constant. In the light-front formulation of quantum field theory, phenomena such as the GMOR relation -- usually  associated with condensates in the instant form vacuum --  are properties of the the hadronic LF wavefunctions themselves.   An exact Bethe-Salpeter analysis~\cite{Chang:2013epa} shows that the quantity that appears in the  Gell Mann-Oakes-Renner relation (GMOR)~\cite{GellMann:1968rz}  is the matrix element $\langle 0|\bar \psi \gamma_5 \psi|\pi \rangle$  for the pion to couple locally  to the vacuum via a pseudoscalar operator  -- not a vacuum expectation value $\langle 0|\bar\psi \psi|0 \rangle$. In the front-form  $\langle 0|\bar \psi \gamma_5 \psi|\pi \rangle$ involves the pion LF Fock state with parallel $q$ and $\bar q$ spin and $L^z= \pm 1$.

In the color-confining light-front holographic model discussed here, the vanishing of the pion mass in the chiral limit, a phenomenon usually ascribed to spontaneous symmetry breaking of the chiral symmetry,  is  obtained specifically from the precise cancellation of the LF kinetic energy and LF potential energy terms for the quadratic confinement potential. This mechanism provides a  viable alternative to the conventional description of nonperturbative QCD based on vacuum condensates, and it 
eliminates a major conflict of hadron physics with the empirical value for the cosmological constant~\cite{Brodsky:2009zd,  Brodsky:2010xf}.

The treatment of the chiral limit in the LF holographic approach to strongly coupled QCD is thus substantially different from the standard approach  based on chiral perturbation theory.
In  conventional discussions,
spontaneous symmetry breaking  by a  non-vanishing chiral quark condensate $\langle  \bar \psi \psi  \rangle$ plays the crucial role. One then expects that the masses of chiral partners with $L$ and $L+1$ become degenerate at high mass. In contrast, the Regge trajectories of states with different $L$ remain parallel in our approach.
Instead, the proton eigensolution itself has $L=0$ and $L=1$ components with equal weight when $m_q=0.$   This observation opens up new possibilities for understanding the role of chiral symmetry and its breakdown in hadron physics.

QCD sum  rules \cite{Shifman:1978bx}  $\langle  \bar \psi \psi \rangle$ involve  non-perturbative elements into the perturbatively calculated spectral sum rules.  It should  be noted, however, that the definition of the condensate, even in lattice QCD necessitates a renormalization procedure for the operator product, and it is not a directly observable quantity.  In contrast, in  Bethe-Salpeter~\cite{Maris:1997hd,Chang:2013epa} and light-front analyses~\cite{Brodsky:2012ku}, the GMOR relation~\cite{GellMann:1968rz}
for $m^2_\pi/m_q$ involves the decay matrix element $\langle 0 |\bar \psi\gamma_5 \psi |\pi \rangle$ instead of $\langle 0| \bar \psi \psi | 0\rangle$.
The GMOR relation thus relates the pseudo-vector and pseudo-scalar decay matrix elements of the pion, not vacuum expectation values. Note also that the LF confining interaction generates a higher twist term of order $\kappa^4\over s^2$ in the annihilation ratio $R_{e^+ e^-}(s)$ which can replace the phenomenological effect of the gluon condensate 
$\langle\alpha_s G^2\rangle$.

\section{Conformal Symmetry in pQCD and The Principle of Maximum Conformality}

The underlying conformal symmetry of the classical chiral QCD Lagrangian  plays a critical role in our analysis.  
Another important application is the ``Principle of Maximum Conformality"~\cite{Brodsky:2013vpa,Brodsky:2011ig} which fixes the renormalization scale at each order of perturbation theory and gives scheme-independent predictions at finite order in perturbative QCD by systematically identifying and resuming the nonconformal $\beta$ terms into the QCD running coupling.  The identification of the $\beta$ terms at every order -- needed for unambiguous BLM/PMC scale-setting~\cite{Brodsky:1982gc} -- can be performed unambiguously using the $R_\delta$ method~\cite{Mojaza:2012mf}. The PMC method satisfies all principles of the renormalization group. It reduces in the Abelian limit to the standard method used for scale-setting for precision QED computations,  sets the number of active flavors $n_F$ correctly at each order, and it can be applied to multi-scale problems. 
PMC scale-setting also eliminates~\cite{Brodsky:2012ik} the conflict between the CDF measurement for the $p \bar p \to t \bar t X$ asymmetry and pQCD  predictions which are based on a guessed scale.  A high precision application of the PMC to Higgs decay is given in Ref. ~\cite{Wang:2013bla}

\section {Future Directions}

The successes of the 
light-front holographic approach for describing the light-quark meson and baryon  spectroscopy  could  be extended in many new directions: 

\begin{enumerate}

\item The LF kinetic energy operator  for massive fields is $\sum_i \frac{k^2_{\perp i} + m_i^2}{x_i}$ summed over the constituents. At first approximation the eigenmass squared is changed by  $\langle \frac{m^2_q}{ x_q}\rangle $
 as in the Feynman-Hellmann theorem.  The effective confining potential can  be modified by replacing $\zeta^2_\perp \to \zeta^2_\perp +\zeta_3^2$ where $\zeta_3= i {\partial \over \partial k_3}$ is conjugate to $k_3 = {2m(x-1/2)\over \sqrt{x(1-x)}}.$

\item
 The  scale-invariant QCD conformal interactions of QCD correspond to short-distance contributions from gluon exchange, leading to hyperfine and spin-orbit splittings in the spectrum. 
The gluonic interactions also lead to sea quark contributions to the Fock state wavefunctions of the hadron as well as BFKL evolution of structure functions and ERBL evolution of the distribution amplitudes.  One thus predicts a two-component theory with harmonic oscillator confinement at long distances and scale invariant gluonic interactions at short distances.  One can see a transition between these components in the behavior of the QCD running coupling constant which behaves as $\alpha_s(Q^2) \propto \exp \left(- Q^2/ 4 \kappa^2 \right)$ for $Q^2 < 4 \kappa^2$ and the logarithmic fall-off dictated by asymptotic freedom for $Q^2> 4 \kappa^2$.   A similar transition can be seen in the pion LF wavefunction determined from the $k_\perp$ fall-off of diffractive dijet production.

\item  Dyson-Schwinger analyses~\cite{Maris:2003vk} show that the light-quarks have effective running masses  $m_q(k^2), $  even if the current quark mass and its Higgs coupling vanishes~\cite{Chang:2013pq}.  In effect AdS/QCD incorporates the dynamical effects of the running masses.   The Dyson-Schwinger analyses also predict that the pion valence LFWF has three $J^z= S^z+L^z=0$  components with $S^z=\pm 1, L^z= \mp 1$  as well as $L^z=0, S_z=0.$  This requires  in a more advanced theory,  a $3 \times 3$ matrix formulation of the pion's valence eigenstate. The AdS/QCD approach for a meson of spin $J$ thus requires a multidimensional LF Hamiltonian matrix which acts on the column vector of valence states  with the set of allowed  orbital angular momenta.  In the case of higher particle-number Fock states, the Born Oppenheimer method could be applied.

\item As shown by Polchinski and Strassler~\cite{Polchinski:2001tt}, AdS/QCD  reproduces the dimensional (quark) counting rules~\cite{Brodsky:1973kr,Matveev:1973ra} for 
hard hadronic exclusive amplitudes and form factors since the leading behavior of the scattering amplitude is controlled by the twist of each hadron's interpolating operator at small $z$. For example, the fall-off of  the LF helicity-conserving cross section 
${d\sigma\over dt} (A+B \to C + D)$ scales as $F(\theta_{CM}) \over s^{n_{tot}-2}$  at large $s$ and  fixed $\theta_{CM}$, where $n_{tot} = n_A+ n_B +n_C +n_D$  is  the total number of active quark or gluonic fields entering the hard subprocess  This leading-twist analysis can extended to predict the angular and spin dependence of such amplitudes as well as the modifications from higher-twist contributions as well as the starting scale for ERBL evolution~\cite{Lepage:1979zb,Efremov:1979qk}.  The AdS/QCD wavefunctions can also be used to compute the quark interchange contributions~\cite{Gunion:1973ex} to exclusive hadronic scattering amplitudes.   This also leads to the prediction that Regge trajectories $\alpha_R(t) $ intercept at negative integers~\cite{Blankenbecler:1973kt} at large space-like momenta $t<0.$ An explanation of why quark interchange amplitudes dominate gluon exchange amplitudes phenomenologically is also possible.  One can also extend the formalism to deeply virtual Compton scattering amplitudes such as $\gamma^* p \to \gamma p$ and $\gamma^* p \to \pi^+ n.$

\item The AdS/QCD model provides a first approximation to the full QCD theory.  In the BFLQ (Basis Light-Front Quantization) approach~\cite{Vary:2009gt}, one constructs an orthonormal basis of confined LFWFs which is then used to diagonalize the full 
Hamiltonian $H_{LF}^{QCD}$. In principle, one can solve the nonperturbative LF QCD Heisenberg eigenvalue problem on this basis.   Since AdS/QCD appears to reproduces much hadron physics, one expects more rapid convergence than the traditional Discretized Light-Cone Quantization ( DLCQ)  procedure, which is based on a free Fock state plane wave basis.

\end{enumerate}

\section{Conclusions}

Light-Front Hamiltonian theory, derived from the quantization of the QCD Lagrangian at fixed light-front time 
$x^+= x^0+x^3 = t + z/c$, provides a rigorous frame-independent framework for solving nonperturbative QCD.    The eigenvalues of the QCD LF Hamiltonian give the invariant mass squared of the hadrons; its eigensolutions are the frame-independent LF wavefunctions which underly hadron phenomenology.   Given the frame-independent light-front wavefunctions (LFWFs) $\psi_{n/H}$, one can compute a large range of hadronic observables, starting with form factors, structure functions, generalized parton distributions, Wigner distributions, etc. Light-front physics is a fully relativistic field theory, but its structure is similar to nonrelativistic atomic physics, and the resulting bound-state equations can be formulated as relativistic Schr\"odinger-like equations at equal light-front time. 
Light-front holography provides a precise relation between the bound-state amplitudes in the fifth dimension of AdS space and the boost-invariant light-front wavefunctions describing the internal structure of hadrons in physical space-time.

One of the most fundamental problems in Quantum Chromodynamics is to understand the origin of the mass scale that controls the range of color confinement and the hadronic spectrum.  For example, if one sets the Higgs couplings to quarks to zero, then no mass parameters appear in the Lagrangian, and the theory is conformal at the classical level.   Nevertheless, the proton and other hadrons have a finite mass. A traditional assumption is that the QCD mass scale of the chiral QCD Lagrangian enters through the mass scale of the renormalized running coupling  ``dimensional transmutation."  However, the analytic mechanism  which leads to quark and gluon confinement and a scheme-independent fundamental mass scale is not apparent. 

An explanation of the origin of the mass scale in nominally conformal theories was given 
by  V.~de Alfaro, S.~Fubini and G.~Furlan~\cite{deAlfaro:1976je} in the context of one-dimensional 
quantum field theory.
It was shown in this paper that
the mass scale which 
breaks dilatation invariance can appear in the equations of motion without violating  the conformal invariance of the action.  In fact, this is only possible if the resulting potential has the form of a confining harmonic oscillator, and the transformed time variable $\tau$ which appears in the confining theory has a limited range.

We have shown that the application of the dAFF procedure,
together with light-front quantum mechanics and light-front holographic mapping,
leads to a new analytic approximation to QCD -- a light-front Hamiltonian and corresponding one-dimensional light-front  Schr\"odinger and Dirac equations which are frame-independent, relativistic, and reproduce crucial features of the spectroscopy and dynamics of the light-quark hadrons. The valence Fock-state wavefunctions 
of the light-front QCD Hamiltonian satisfy a single-variable relativistic equation of motion, analogous to the nonrelativistic radial Schr\"odinger equation, with an effective 
confining potential $U_{\rm eff}$ which systematically incorporates the effects of higher quark and gluon Fock states.   The confining term in the LF potential has a unique form of a harmonic oscillator in the impact variable $\zeta^2$ if one requires that the  action remains conformally invariant.  The holographic mapping of gravity in AdS space to QCD with the specific ``soft-wall" dilaton $e^{\kappa^2 z^2}$ yields the same confining potential $U(\zeta^2)$ which is consistent with conformal invariance of the QCD action and the 
light-front Schr\"odinger equation, but with an extra term $2 \kappa^2(J-1)$
which is derived from the embedding space when extended to hadrons with spin $J$. 

We thus obtain an effective light-front effective theory for general spin which 
encodes the conformal symmetry of the four-dimensional classical QCD Lagrangian in the form of a 
light-front Schr\"odinger equation.
The predictions include a zero-mass pion in the chiral $m_q\to 0$ limit, and linear Regge trajectories $M^2(n, L, S) = 4\kappa^2( n+L +S/2)$ with the same slope  in the radial quantum number $n$ and the orbital angular momentum $L$. The corresponding light-front Dirac equation provides a dynamical and spectroscopic model of nucleons.  We have also outlined a number of interesting future applications. 

This 
light-front holographic approach predicts not only hadron spectroscopy  successfully, but also hadron dynamics -- hadronic form factors, the QCD running coupling at small virtuality, the light-front wavefunctions of hadrons, $\rho$ electroproduction, distribution amplitudes, valence structure functions, etc. 
Thus  the combination of 
light-front dynamics, its holographic mapping to gravity in a higher dimensional space, and the dAFF procedure 
provides new  insights into the physics underlying color confinement, chiral invariance, and the origin of the QCD mass scale, among the most profound questions in physics.

\section*{Acknowledgments}

Presented by SJB at  the International Conference on Flavor Physics and Mass Generation, Singapore, February 10-14, 2014.  
This research was supported by the Department of Energy, contract DE--AC02--76SF00515.  SLAC-PUB-15937.
We thank Arek Trawinski, Stan Glazek and our other collaborators for helpful conversations.

{}

\end{document}